\documentclass[referee]{aa} 

\usepackage{graphicx,amssymb,psfig,natbib}

\newcommand{\AaA}{A\&A}

\newcommand{\ApJS}{Astrophysical Journal Supplement Series}
\newcommand{\AJ}{AJ}
\newcommand{\ApJ}{ApJ}
\newcommand{\MNRAS}{MNRAS}

\newcommand{\Msun} {M$_\odot$}
\newcommand{\Lsun} {L$_\odot$}
\newcommand{\Teff} {T$_{\rm{eff}}$}
\newcommand{\Tstar} {T$_{\rm{eff}}$}
\newcommand{\Lstar} {L$_\star$}
\newcommand{\Mstar} {M$_\star$}

\newcommand{\um} {$\mu$m}

\newcommand{\Md} {M$_{d}$}
\newcommand{\Ri} {R$_{in}$}
\newcommand{\Rd} {R$_{d}$}

\newcommand{\simless}{\mathbin{\lower 3pt\hbox
      {$\rlap{\raise 5pt\hbox{$\char'074$}}\mathchar"7218$}}} 
\newcommand{\simgreat}{\mathbin{\lower 3pt\hbox
     {$\rlap{\raise 5pt\hbox{$\char'076$}}\mathchar"7218$}}} 

\begin{document}

\title{Spatially resolved PAH emission in the inner disks of Herbig Ae/Be stars}

\author{Emilie Habart\inst{1},
Antonella Natta\inst{1}, 
Leonardo Testi\inst{1}, 
Marcel Carbillet\inst{2}}

\institute{ 
    Osservatorio Astrofisico di Arcetri, INAF, Largo E.Fermi 5,
    I-50125 Firenze, Italy 
\and
    Laboratoire Universitaire d'Astrophysique de Nice, UMR 6525 du CNRS, Parc Valrose, 06108 Nice Cedex 2, France}

\abstract{We present adaptive optics high angular resolution ($\sim$0.1$\arcsec$) spectroscopic 
observations in the 3~\um\ region of eight well known Herbig Ae/Be stars with circumstellar disks. 
We detect the aromatic emission feature at 3.3~\um\ for four out of six of our objects with flared 
disks (HD~169142, HD~97048, HD~100453, HD~100546), some weaker additional features at 3.4 and 
3.46~$\mu$m and nanodiamond features at 3.43 and 3.53~\um\  in two of our flared object 
(HD~100546 and HD~97048 respectively).  We also detect hydrogen recombination line at 3.74~\um\ in 
practically all objects.
The emission in the polycyclic aromatic hydrocarbons (PAHs) feature at  3.3~\um, additional and nanodiamond features in the 3.4-3.5~\um\ region 
is, for the first time, spatially resolved in all the sources where the features are detected.
The full-width at half-maximum sizes that we derive are typical of emission arising in a circumstellar 
disk. On the other hand, the continuum emission is unresolved, with the exception
of HD~97048 where it is marginally resolved.
We compare the observed spatial distribution of the 3.3~$\mu$m PAH feature and the adjacent continuum 
to the predictions of a disk model that includes transiently heated small grains and PAHs in addition 
to large grains in thermal equilibrium \cite[]{habart2004a}. The model predicts that, as observed, 
the 3.3~$\mu$m PAH emission feature is significantly broader than that of the adjacent continuum
and that about 50\% of its integrated intensity comes from a radius $R<$ 30 AU.
We find that the predicted brightness profiles reproduce very well the observed ones.
This proves beyond doubt that the energetic 3.3~$\mu$m PAH emission feature takes its origin in 
the inner disk regions.
 }

\authorrunning{Habart et al.}
\titlerunning{PAH emission in the inner disks}

\maketitle

\section{Introduction}
Aromatic infrared emission bands (AIBs) attributed to PAHs have been detected in a number of pre-main sequence star of intermediate mass. According  to the more recent statistics based on ISO spectra, almost 
60\% of the Herbig Ae/Be  stars (hereafter HAeBe) show PAHs in emission \cite[]{acke2004}.

In \cite{habart2004a}, we have investigated the presence and properties of PAHs in circumstellar disks of HAeBe stars by comparing the predictions of disk models to observations with ISO and ground-based telescopes. We show that PAHs are present at the surface of disks,  and absorb a significant fraction of the stellar radiation, similarly to PAHs in the interstellar medium. Consequently, PAHs are an important source of opacity in circumstellar disks and are likely to play a dominant role in the thermal budget (via the photoelectric effect) and chemistry of the gas \cite[via H$_2$ formation on dust surfaces,][]{habart2004b}. 
Moreover, the PAH emission is predicted to be extended, more that than of 
large grains at similar wavelengths.
This is easily understood when considering the different
excitation mechanism of the small transiently heated particles and of the large
grains at thermal equilibrium.  
Because of the extension of the emission, which can be more easily resolved than that of the adjacent continuum with state-of-the-art adaptive optics (AO) systems and
interferometers, one can use the small particles emission features to
probe  disk properties, such as the inclination on the line of sight
of the inner disk. 

Recently, using AO high angular resolution spectroscopic observations of the HAeBe star HD~97048, we were able for the first
time to spatially resolve the emission
in the nanodiamond features at 3.43 and 3.53~$\mu$m \cite[]{habart2004c}. 
We find that the diamond emission originates in the inner region
($R \lesssim 15$ AU) of the disk and show that the disk is seen pole-on, confirming previous indirect evidence.

Here we present high angular resolution (0.1$\arcsec$) spectroscopic observations
in the 3~\um\ region obtained with NAOS-CONICA at the VLT of a sample of HAeBe stars. These observations allow us to spatially resolve 
the inner disk emission in the PAH emission feature at 3.3~\um\ due to C-H bond stretching vibrations, as well as 
in some additional and nanodiamond features in the
3.4-3.5~\um\ region.
The paper is organized as follows. 
Sect. \ref{sample} presents our sample.
Sect. \ref{observation} describes the observations and the data reduction.  
The results  are presented in Sect \ref{result}. In Sect. \ref{model}, we compare the observations to the model predictions. Sect. \ref{summary} summarizes our  results.

\section{The sample}
\label{sample}

Our sample consists of 8 well known HAeBe stars, all with evidence of
a circumstellar disk
and for which there are many observations at different wavelengths,
including ISO spectroscopy. 
In Table \ref{table_sample}, we report (i) the astrophysical parameters of each star (distance,
spectral type, effective temperature, luminosity, and $\chi$, i.e., far-ultraviolet (FUV) flux at 10 AU from the
star), (ii) the observational characteristics of the disk (evidence for a flared
disk) and (iii) the dust spectral characteristics
(presence of AIBs and silicate bands). 
We also give the strength of the 3.3~\um\ aromatic emission feature 
 measured from the ISO spectra of \cite{meeus2001} and \cite{vankerckhoven2002}.

In this sample, we can distinguish three different groups \cite[]{meeus2001}: 
the first (group Ia) is made of stars with evidence of
a flared disk and strong or moderate aromatic and silicate bands; the second (group Ib) is also characterised by
flared disks and the presence of AIBs, but the silicate bands are absent; finally,
in the third group (group II) there is no evidence of a flared disk, there
are weak or no
AIBs, but the silicate emission features are strong.

\section{Observations and data reduction}
\label{observation}

Long slits spectra in the L-band (3.2--3.76~\um) were taken with the
adaptive optics system NAOS-CONICA (NACO) at the VLT, using the visible 14$\times$14 Shack-Hartmann wave-front
sensor working at nearly 500\,Hz. The 28$\arcsec$ long slit had a width of 0$\farcs$086 which roughly corresponds
to the diffraction limit while the pixel scale was $\sim$0$\farcs$0547
and the spectral resolution $\sim$700.
For all objects, we took one slit position centered on the star.
For one of our object, HD~97048, we took 7 slit positions: one centered on the star and the others at 0$\farcs$043, 0$\farcs$086, 0$\farcs$129 and -0$\farcs$043, -0$\farcs$086, -0$\farcs$129 off axis \cite[see also][]{habart2004c}.
The long slit was aligned in the north-south direction, except for HD~100546.
For HD~100546, the long slit was aligned with the major axis of the disk resolved
in scattered light \cite[]{augereau2001,grady2001b}, with a position angle of $\sim$160$^{\circ}$ measured east of north.
For each object, in order to correct from the atmospheric and instrumental background, we employed
standard chop/nod technique with a throw of $\sim$9\arcsec\ in the north-south direction.
 The integration time per chop- or nod-positions was 1~min and the total integration time per source was about 30 min.

A summary of the observations is given in Table \ref{table_obs}. 
The log of the observational conditions reports Fried parameters $r_0$
ranging from 7 to 25\,cm (at 500\,nm), average outer-scale ${\cal L}_0$ from 13 to 23 m, and a resulting average coherent energy (after AO correction and at 2.2~\um) from 25 to 60\% on axis. The latter roughly gives the
actual Strehl ratio characterizing the quality of our data. According to 
simultaneous measurements, the  seeing (at 500\,nm) was between 0.6 and 1.5$\arcsec$ consistent
with
the NACO values. Our observations clearly benefit from good seeing conditions and AO
correction, and the achieved angular resolution was close to the diffraction limit.

Data reduction was performed using a standard procedure for IR spectroscopic observations.
To remove the telluric features we used the observations of spectroscopic
standard stars (listed in Table \ref{table_obs}) taken for all sources  immediately before or after the science target
\cite[except HD~97048, see][]{habart2004c}.
These references stars providing a good match to the objects in terms of atmospheric turbulence conditions ($r_0$, ${\cal L}_0$,
seeing) and AO correction quality were also used to estimate the point-spread function (PSF).

\section{Results}
\label{result}

\subsection{Spectra}
\label{spectra}

Fig. \ref{spectra} shows the spectra obtained for the sources in our sample integrated over the slit for  distances from the star  $\leq$1\arcsec. 
When available, the figure also shows for comparison
the spectra obtained with the Short Wavelength Spectrometer (SWS) on board ISO with a beam of $\sim$14$\times$20$\arcsec$ around 3~$\mu$m
and published in \cite{meeus2001} and \cite{vankerckhoven2002}.
Both NACO and ISO spectra are normalized to the continuum emission adjacent  to
the 3.3~\um\ PAH feature. 
The wavelength coverage of the NACO spectra allows us to observe the 3.3~\um\ PAH feature, some additional and nanodiamond features
in the 3.4-3.5~$\mu$m region, and two hydrogen recombination lines at 3.297 and 3.741~$\mu$m, as discussed in the
following.

\subsubsection{The 3.3~\um\ PAH feature}

In our NACO spectra, we clearly detect the 3.3~\um\ PAH feature  for four out of six of our objects of group I, whereas we do not for the two objects of group II. 

For the foor group I objects showing the 3.3~\um\ feature (HD~169142, HD~97048, HD~100453, HD~100546), the feature is detected by both NACO and ISO. 
Comparing the NACO and ISO spectra, we find that the ratio of the emission peak to the continuum is roughly similar, i.e., that the larger ISO beam does not add more emission in the
feature. 
This confirms that the PAH emission is confined to the disk. 
The  exception is the ISO spectrum for HD~97048 which shows
stronger PAH emission at 3.3~$\mu$m than the NACO one. In this case,
one can think that the surrounding nebula contributes to the emission
\cite[e.g.,][]{boekel03}.

The two group I objects where the 3.3~\um\ feature
is not detected are AB Aur and CQ Tau.
In the case of AB Aur, the 3.3~\um\ feature
is not detected by neither NACO nor ISO.
On the other hand, PAH features at longer wavelengths (i.e., 6.2, 7.7, 8.6~\um) 
have been detected  \cite[]{vankerckhoven2002,acke2004}. 
One possible explanation of the non detection of the 3.3~\um\ 
feature is that in this object, which presents a particular strong near-IR continuum emission, the 3.3~\um\ feature disappears in contrast to the continuum,
but other possibilities cannot be excluded, including a dominant
contribution to the longer wavelength features from a residual envelope.

For CQ Tau, no ISO-SWS spectra around 3~$\mu$m are available. 
In this system, the non detection of the 3.3~\um\ 
feature may be due to low excitation conditions due to the later spectral type (F2) of the central star.

In the group II objects (HD~142666, HD~163296), an interesting case is HD~142666 which shows in the ISO spectra weak 3.3~\um\ feature and clear evidence of 
the 6.2~\um\ feature.
In the NACO data, 
the signal-to-noise ratio is not sufficient to claim the presence or absence of a weak 3.3~\um\ feature.
For HD~163296, 
there is certainly no strong 3.3~\um\ feature but the data were noisy 
and we cannot say anything about a weak feature. 

In general, the NACO spectra
indicate that PAH emission has size typical of disk and that the PAH emission is strong when the disk is flared, as suggested by previous studies based on ISO data \cite[e.g.,][]{meeus2001,vankerckhoven2002,acke2004,habart2004a}.

\subsubsection{The features in the 3.4-3.5~\um\ region}
\label{satellite}

In two  objects (HD~97048, HD~100546)
the NACO spectra show, in addition to the aromatic band at 3.3~\um,
some features in the 3.4-3.5~\um\ region.

HD~97048 shows the peculiar strong features peaking at 3.43 and 3.53~\um, 
attributed to nanodiamond surface C-H stretching modes \cite[e.g.,][]{guillois99,vankerckhoven2002a,sheu2002,jones2004}. 
As discussed in \cite{habart2004c}, we see that the diamond components are in the NACO and ISO spectrum very similar in peak position, width and
relative strength. Moreover, the peak/continuum ratio in the diamond bands is roughly similar.
This suggests that nanodiamond emission is confined to the disk.

HD~100546 shows different, weaker features, peaking at 3.4 and 3.46~\um.
In the NACO data, these features are clearly visible for distances from the star from $\sim$0.2$\arcsec$ to 0.6$\arcsec$ (see Fig. \ref{profiles_add_bis}).
In the ISO data, the shape of these bands is uncertain since the up and down scans differ \cite[]{vankerckhoven2002}. 
These weak ``satellite'' bands have been observed in several PAHs sources, including 
reflection nebulae, planetary or proto-planetary nebulae \cite[e.g.,][]{geballe85,jourdaindemuizon90,jourdaindemuizon90a,joblin96,vankerckhoven2002}. 
They could be attributed to aliphatic C-H stretches in methyl or ethyl side-groups attached to PAHs \cite[e.g.,][]{joblin96}.

\subsubsection{The hydrogen emission lines}
\label{hydrogen}

Finally, the observed spectral interval contains two
hydrogen emission lines, namely Pf~$\delta$ close to the center of the broad PAH feature at 3.3~\um\ and Pf~$\gamma$ at 3.74~\um. 
Pf~$\gamma$ is visible in practically all objects and is
stronger in the smaller aperture NACO data than in the ISO spectrum; its emission come mostly from the
innermost regions of the  system (see Sect. \ref{profile_hydrogen}).

In the following, we use the observed Pf~$\gamma$ intensity
to estimate the contribution of Pf~$\delta$ to the 3.3~\um\ feature,
using the theoretical intensity ratios given by \cite{hummer87}.
For most of our objects, we find that the contribution of Pf~$\delta$ to the 3.3~\um\ feature is negligible ($\le$4\%).
The exception is HD~100546 where Pf~$\gamma$ is strong and Pf~$\delta$ seriously affects the observed shape of the 3.3~\um\ feature. 
For this object, we find that the contribution of Pf~$\delta$ to the 3.3~\um\ feature 
is strong close to the star
and decreases very rapidly in the outer parts.
For distances from the star $\le$0.1$\arcsec$, the spectrum at 3.3~\um\ is completely dominated by the contribution of Pf~$\delta$, we do not see the PAH emission and the measured Pf~$\gamma$/Pf~$\delta$ ratio is similar to that predicted by \cite{hummer87}.  In the NACO spectrum for distances from the star $\le$ 1$\arcsec$, the contribution of Pf~$\delta$ to the 3.3~\um\ feature is about 40\% ; this contribution is lower in the ISO data, about $\lesssim$7\%, 
and the observed shape of
the 3.3~\um\ feature is less affected.

\subsection{Brightness spatial distribution}
\label{profiles}

In this section, we present the brightness spatial distribution
along the slit for the 3.3~\um\ aromatic feature, the features in the 3.4-3.5~\um\ region and the HI line at 3.74~\um.
We also show the brightness spatial distribution of the continuum 
emission adjacent to the 3.3~\um\ PAH feature, which is due to warm large grains emission from the disk and/or stellar photospheric emission, as discussed in the following and in Sect. \ref{model}.

\subsubsection{Spatially extended 3.3~\um\ PAH feature}
\label{profiles_PAH}

Fig.~\ref{profiles} shows 
the intensity profile along the slit  of the 3.3~\um\ PAH feature, the adjacent continuum and  the observed reference star (the assumed PSF) for the 4 sources where we detect  the 3.3~\um\ PAH feature. 
The 3.3~\um\ emission feature has been  corrected for
the contribution of Pf~$\delta$,  as described in Sect. \ref{hydrogen}.
Errors on the measured full-width at half-maximum (FWHM) size values are dominated by systematic errors at the level of 20\%.

In all objects
the spatial extent of the PAH emission is significantly broader
than that of the PSF. On the other hand, the  continuum emission is unresolved, with the exception
of HD~97048 where it is marginally resolved \cite[see also][]{habart2004c}.

\par\bigskip\noindent
{\bf HD~169142.} The PAH emission feature at 3.3~\um\  
is extended with a FWHM of 0.3$\arcsec$ (or 43 AU).
The feature is clearly visible above the continuum and is strong in the inner regions (radius $\le$0.3$\arcsec$), decreasing rapidly in the outer regions.  
For radius $>$0.3$\arcsec$, the feature is weak and 
it is not possible to measure its brightness profile with sufficient accuracy.

\par\bigskip\noindent
{\bf HD~97048.} The 3.3~\um\ emission feature appears 
extended as in HD~169142.
On one section of the slit and for distances from the star from 0 to 0.2$\arcsec$, the NACO spectrum was noisy near the 3.3~\um\ feature and it was not possible to measure its brightness profile with sufficient accuracy.
As in HD~169142, the 3.3~\um\  feature is weak in the outer regions and given the uncertainties of the data,  differences in the wings between the two objects are not significant.
The continuum emission is spatially resolved in HD~97048 with a FHWM of 0.180$\arcsec$ (or 32~AU); this indicates that in this system 
the continuum emission is mostly due to warm large grains emission from the disk.
For HD~97048 for which we obtained NACO spectra on several slit positions, we were able to reconstruct two-dimensional images of the inner disk in both
the continuum emission and the nanodiamond emission features \cite[see Fig. 4,][]{habart2004c}. The roughly circular shape of the emission maps confirms the hypothesis of a pole-on disk in HD~97049 put forward by \cite{the86}.

\par\bigskip\noindent
{\bf HD~100453.} The 3.3~\um\ feature emission  is marginally resolved, with
a FWHM of 0.18$\arcsec$ (or 20 AU). 
In this object, the PAH emission feature decreases very rapidly in the outer zones. This could be due to low excitation conditions due to the later type (A9) of the central star.

\par\bigskip\noindent
{\bf HD~100546.} The PAH emission spatial distribution in HD~100546 differs in several points from that found in the other objects.  
Firstly, we detect the presence of a gap in the innermost region ($R\le$ 0.05-0.1$\arcsec$ or 5-10 AU).
In this region, as discussed in Sect. \ref{hydrogen}, the spectrum at 3.3~\um\ is in fact completely dominated by the contribution of Pf~$\delta$ and we do not see the PAH emission. 
This gap is also found for the emission of the additional weaker features in the 3.4-3.5~\um\ region (see Sect. \ref{profile_satellite}) where there are no HI lines. 
It could be due to the lack of the carrier of the bands in the regions very near the star or to the fact that the emission of these features disappears completely in contrast to the continuum. But it could also result from the presence of a ``physical hole''  
as suggested by the analysis of the ISO spectra of HD~100546 by \cite{bouwman2003}.  In order to explain the amount of mid-IR dust emission and the observed crystalline features, \cite{bouwman2003} propose that the disk should 
have a hole and a wall at about 10 AU.
In the NACO data there is no gap in the 3~\um\ continuum emission,
as expected in the case of a ``physical hole''.
However, we cannot rule out, on the basis of our data, that
the stellar photosphere contributes a major fraction of the 3~$\mu$m flux.
Higher spatial resolution data combined with an accurate analysis of the
spectral energy distribution (SED) are necessary to shed light on this very interesting object. 

We also find that the 3.3~\um\ emission feature is at larger distances stronger in HD~100546 than in the other objects. We can easily trace  the 3.3~\um\ emission feature to distances of $\sim 0.5$ arcsec or 50 AU from the star. Because of the presence of the gap, it is not possible to determine its FWHM size, but we note that our NACO data are consistent with a FWHM of $\sim$0.2$\arcsec$ (20 AU) 
recently measured from ISAAC spectroscopic observations \cite[]{geers04}.

\subsubsection{Spatial extension of the features in the 3.4-3.5~\um\ region}
\label{profile_satellite}

We now look at the spatial emission distribution of the features in the 3.4-3.5~\um\ region observed in two of our objects (HD~97048, HD~100546) and compare it to that of the 3.3~\um\ PAH feature.

Fig. \ref{profiles_add} shows the intensity profile of the strongest feature at 3.53~\um\, attributed to diamonds, observed in HD~97048.
The intensity profile of the other strong diamond feature at 3.43~\um\ is similar. The derived FWHM is of 0.23$\arcsec$ (41 AU) in both the 3.43 and 3.53~\um\ features \cite[see][]{habart2004c}. 
Also, by comparing the spatial extension of these features to that of the continuum and of the PAH at 3.3~\um, we find that the diamond emission is extended more than the continuum 
 but less  than that of the PAHs.
This is roughly consistent with the idea that the diamond carrier of the 3~\um\ features
observed in HD~97048 have sizes intermediate between those of large grains and of PAHs,
 as suggested by detailed analysis of the ISO spectrum  \cite[see e.g.][]{vankerckhoven2002a,sheu2002,jones2004}.
Finally, we note that recent AO observations at the Subaru telescope of Elias 1, which shows 
the same diamond bands at 3.43 and 3.53~\um\ as HD~97048, show similar results (M. Goto, private communications).
 The observed FWHM of the
diamond features is similar to that derived here for HD~97048
and the PAH emission is also found to be spatially more extended than that of the diamonds.

Fig. \ref{profiles_add_bis} shows the intensity profile of the strongest additional feature at 3.4~\um\ observed in HD~100546. 
As for the PAH 3.3~\um\ feature, there is evidence of a gap in the innermost regions.
Also,  the emission of the ``satellite'' feature is  extended as the PAH
emission. This is roughly consistent with the idea that these features are due to aliphatic C-H stretches in methyl or ethyl side-groups attached to PAHs.

\subsubsection{Spatial extension of the hydrogen emission lines}
\label{profile_hydrogen}

Finally, we look at the spatial extension of the HI Pf~$\gamma$ recombination 
line at 3.74~\um. Fig. \ref{profiles_HI} shows the intensity profile of Pf~$\gamma$ observed in two of our objects (HD~169142, HD~100546) where we clearly 
detected it. 
The HI emission line is not
resolved. It originates from ionized material very close to the star,
possibly  the ionized gaseous disk or matter accreting from the inner disk onto the star.

\section{Comparison to model predictions}
\label{model}

In this section, we compare the observed spatial distribution of the 3.3~$\mu$m PAH feature to the predictions of a disk model that includes PAHs and very small grains in addition to large grains. These models \cite[]{habart2004a} 
reproduce the strength of the most commonly observed PAH features (i.e., 3.3, 6.2, 7.7, 11.3~\um) in many HAeBe stars.

\subsection{Description of the model}
\label{description}

In \cite{habart2004a}, we have computed models
of disks heated by irradiation from the central star, in hydrostatic equilibrium in the vertical
direction, with dust and gas well mixed (flared disks).
We have used a 2-layer model to calculate the disk structure
and implemented a 1D radiative transfer code 
to compute the emerging spectrum. 

The models are appropriate for disks that are optically thick to the stellar
radiation. This is generally the case for disks around pre-main--sequence stars,
up to very large radii. 
For simplicity, we limit ourselves here to a disk with \Md$\sim$ 0.1 \Msun,
an inner radius of \Ri = 0.3 AU (close to
the BG evaporation radius) and an outer radius of \Rd=300 AU. The surface density
profile is $\Sigma=\Sigma _0 \times (r/R_i)^{-1}$ with $\Sigma _0$= 2 10$^3$ g cm$^{-2}$.

The dust is a mixture of large grains in thermal
equilibrium, transiently heated small grains and PAHs.
The big grains consist of graphite and silicates with optical constants from
\cite{draine85}; they have an MRN size distribution ($n(a) \propto a^{-3.5}$) between a minimum and a maximum radius $a_l=$0.01~$\mu$m and $a_u=$0.74~$\mu$m for silicates and $a_l=$0.01~$\mu$m and $a_u=$0.36~$\mu$m for graphite. 
To this grain population, we add a tail of the very small graphite particles with the same size distribution ($n(a) \propto a^{-3.5}$) between $a_l=$10 \AA\ and $a_u=$100 \AA. Optical constants are from \cite{draine85}.

For the population of PAHs, the absorption cross section is from \cite{li2001a}. 
In the ISM, PAHs are made up of a few tens up to a few
hundreds of carbon atoms; for reasons of simplicity, we take in our model one PAH size, $N_C=100$. The hydrogen to carbon ratio is $H/C = f_H \times(6/N_C)^{0.5}$ \cite[case of compact symmetric PAHs, see][]{omont86} 
with $f_H$ - the hydrogenation fraction of the molecule -  equal to 1 (i.e., essentially fully 
hydrogenated PAHs). With respect to the charge,
to keep the model simple we neglect that probably we have in disks 
a mixture of neutral and charged PAHs
and assume that PAHs are neutral.
Finally, we take into accont that in a strong FUV radiation field, photo--destruction 
of PAHs will occur during multiphoton events where PAHs absorb an energy of more than 21 eV in a time shorter than the cooling time.
In our model, we assume that PAHs are destroyed when the probability
that they are at a temperature $T\ge T_{evap}$ ($T_{evap}$=2000 K) exceeds a critical value 
$p_{evap}$ ($p_{evap}=10^{-8}$) \cite[for more details see][]{habart2004a}.
The effect of these assumptions is briefly discussed in Sect. \ref{discussion}.

Finally, the silicate abundance in dust is $[Si/H]=3\times 10^{-5}$, and the total carbon
abundance in dust is $[C/H]=2.22\times 10^{-4}$.  Of this, 23\% are in PAHs, 10\% in
VSGs and 67\% in large grains. 

The model predicts an infrared spectral energy distribution showing PAH features at 3.3, 6.2, 7.7, and 11.3~\um\ clearly visible above the continuum emission mostly due to warm large grains \cite[see Fig. 3 in][]{habart2004a}.
In \cite{habart2004a}, we have compared the model results to observations
from ISO and ground-based telescopes of some thirty HAeBe stars, including all our targets except CQ Tau. 
Taking into account the uncertainty due to our disk description and 
assumed dust model, we concluded that
the model can account for the strength of the most commonly observed 
PAH features (i.e., 3.3, 6.2, 7.7, 11.3~\um) in most of the objects with flared disks. 

For geometrically flat disks, PAH features are predicted to be very weak, even when PAHs with standard properties are present on the disk surface.
This agrees with the obervations showing that in objects with no evidence of a flared disk (group II) no (or weak) PAH features have been detected.

\subsection{Predicted spatial distribution}
\label{predicted_profiles}

The model predicts that the PAH emission is much more spatially extended than the adjacent continuum emission, as shown in Fig. \ref{profile_mod}.
Fig. \ref{profile_mod}(a) plots as a function  of the radius the 
cummulative intensity of the PAH feature and continuum
emission at 3.3~$\mu$m. The continuum reachs 50\% of its intensity
at a very small radius (about 2 AU) while the feature does so
at larger radii (about 30 AU).
 This behaviour basically reflects the different excitation mechanism of PAHs, which are transiently heated by individual UV photons, and of the continuum, which is  dominated by the 
emission of big grains, in thermal equilibrium with the radiation
field. Only  big grains located very near the star are hot enough to
 emit at 3~\um, whereas also PAHs  farther away  can be excited
and emit in the 3.3~\um\ feature.
However, also the 3.3~\um\ feature is expected to be stronger 
near the star since the emission of PAHs roughly scales with the intensity
of the FUV radiation field. Moreover, among the PAH features, the 3.3~$\mu$m one is the least extended spatially.
As discussed in \cite{habart2004a}, also for transiently heated particles 
the features at short wavelengths are in fact stronger in
the inner part of the disk (where PAHs are hotter because multiphoton events occur where the radiation field is highest), decreasing rapidly in the outer cold part.
On the other hand, the features at longer wavelengths are 
more extended.  

Fig. \ref{profile_mod}(b) shows the surface brightness profiles 
of the 3.3~\um\ feature and the adjacent continuum.
As expected from Fig. \ref{profile_mod}(a), the PAH emission feature 
is much more extended that the adjacent continuum. 
Finally, it is interesting to note that for the formalism adopted here, we find that
very near the star ($R<$1-2 AU) PAHs are mostly photoevaporated,
as shown by the inner gap seen in the 3.3~\um\ feature continuum subtraced  
emission profile (see Fig. \ref{profile_mod}(b)).
At $R\sim$15 AU, about 50\% of PAHs 
are photo-evaporated and for $R>20$ AU this fraction goes to zero.

\subsection{Comparison with the observations}
\label{comparison_profiles}

Fig. \ref{comparison_mod_obs} shows for each of our objects showing the 3.3~$\mu$m feature
the intensity profile of the 3.3~\um\ feature (continuum subtracted)
obtained by convolving the computed intensity with the corresponding observed PSF. For each object, we have computed a model with the suitable stellar parameters given in Table \ref{table_sample}. The disk and dust parameters are
the same as in the previous discussed model shown in Fig. \ref{profile_mod} and as described in Sect. \ref{description}. The star/disk system is assumed to be seen face-on at the distance of the object.
Fig. \ref{comparison_mod_obs} also shows the predicted intensity profile of the adjacent continuum
 which is the sum of the computed disk continuum emission (mostly due to warm large grains) and of the photosphere stellar emission at 3~\um, each convolved with the observed PSF. The stellar contribution to the continuum emission at 
3~\um\ is typically about 30\% for our stars
with spectral type of B9-A5, while it dominates for the later spectra type stars. The exact value of the stellar contribution depends not only
on the stellar properties but also on disk parameters. Note, however, that the continuuum subtracted intensity profile of the PAH emission is independent of them.

By comparing the predictions to the observations, we find that, as observed,
 the predicted 3.3~\um\ PAH emission is  significantly broader than that of the adjacent continuum and its predicted  
FWHM is very close to the observed one (differences $\le$ 20\%, see Fig. \ref{comparison_mod_obs}).
For some objects, there are intringuing differences between model predictions and observations in the wings, which are however generally within the uncertainties of the data. 
On the contrary, the adjacent continuum emission is not resolved
or is only slighlty broader than the PSF. 
This also agrees with the observations.
We take the general agreement between observations and predictions as a
strong support for the physical pictures underlying our model. 

It must be emphasized that there are several complications which we have neglected. The most obvious is that we have assumed that PAHs can be characterized by a single size, hydrogenation and charge state. This is unlikely to be the case  and is discussed in the next section. However, as discussed in the following,
the predicted brightness profiles of the 3.3~\um\ feature for different PAHs properties are similar in the inner regions  and show only significant differences at large radii, $R\gtrsim 40$ AU.

\subsection{Discussion}
\label{discussion}

Our results allow us to locate the PAH emission  in the disk, to show that the
emission of the PAHs is spatially more extended than that of adjacent continuum and to
prove  beyond doubts that  the energetic 3.3~\um\ PAH feature takes its origin in the inner disk regions.
It is interesting to note that the spatial extension of the 3.3~$\mu$m feature
is much smaller than those observed for PAH features at longer
wavelength
(i.e., 8.6, 11.3, 12.7~\um), which are   
extended on a scale of (several) 100 AU in HD~97048 and HD~100546 \cite[]{boekel03}.
This is consistent with the model results, which showed how,
among the PAH features in disks, the 3.3~$\mu$m one is the least extended spatially (see Sect. \ref{predicted_profiles}).

Our results have the interesting implication that PAHs survive photo-evaporation or coagulation processes 
even if these processes are in principle important in the inner disk,  
where the UV flux and the gas density are very high.
Nevertheless, processes such as photo-evaporation must significantly affect the abundance of PAHs.
For the formalism adopted here, we find that in the $R<20$ AU region
about 60\% of PAHs (with $N_C=100$) are photo-evaporated. 
But PAH evaporation is a complex process which increases rapidly with $\chi$ and for smaller PAHs 
and it is difficult to make quantitative estimates.
Alternatively,
it is  possible that continuous replenishment of PAHs via sublimation of icy mantles (in which interstellar PAHs may have condensed during the dense molecular cloud phase) or by accreting carbon atoms and/or ions from the gas \cite[see, e.g.,][]{allain96} maintain the PAHs population throughout the disk. 
This is a very interesting point 
and future detailed comparison with theoretical calculations will certainly add important information to the studies of  the processes affecting the 
size of grains  in disks.

In the inner disk regions,
processes such as ionization or 
de-hydrogenation are also expected to be important (because of the strong UV field) and to modify the PAHs properties. One  expect PAH properties to vary 
as a function of the radius and depth in the disk. For example, PAHs are likely to be more positively ionized in the inner disk regions. 
Moreover, processes such as photo-evaporation or coagulation could affect the size of PAHs.
Therefore, it is unlikely that PAHs are characterized by a single charge state, size or hydrogenation parameter, as assumed in our model. 
In order to get some insight into the specific
PAH properties, one needs detailed comparison between the observations,
not only of the intensity of the various features, as in \cite{habart2004a},
but also of the brightness profiles at large radii, $R\gtrsim 40$ AU, where
different PAHs have very different emission \cite[e.g., Figs. 6 and 7 in][]{habart2004a}.  
Unfortunatly, our NACO data in the wings are not of sufficient quality
to be used in this context,
but we want to emphasize that future higher sensitive spatially resolved spectroscopic data at $\sim$3~\um\ 
have the potential capability of constraining the PAH properties and of
studying their evolution.

\section{Summary and conclusion}
\label{summary}

We have presented NACO/VLT spectroscopic observations in the 3~$\mu$m region of a sample of HAeBe stars.
These data allow us, for the first time, to spatially resolve the inner disk emission in the bands of small transiently heated particles
and thus provide a benchmark for testing the dust properties within the disk but also the inner disk properties.
Our main results can be summarized as follows.

\begin{enumerate}

\item We spatially resolve the  3.3~\um\ PAH emission for all the sources where this feature is detected.
We find that the PAH emission is significantly broader than that of the adjacent continuum. On the other hand, the continuum emission is unresolved, with the exception
of HD~97048 where it is marginally resolved.

\item We spatially resolve the emission in the nanodiamond features at 3.43 and 3.53~\um\  in HD~97048 and 
find that 
the diamond emission is extended more than the continuum emission but less than the PAH emission. This is roughly consistent with the idea that the diamond carrier of the 3~\um\ features observed in HD~97048 have sizes intermediate between those of large grains and of PAHs, as suggested by detailed analysis of the ISO spectrum  \cite[]{vankerckhoven2002a,sheu2002,jones2004}. 
We also detect weaker additional features at 3.4 and 3.46~\um\ in HD~100546, 
and find their emission is extended as the PAH emission.

\item In HD~100546, we detect the presence of a gap in the innermost regions ($R\le$ 0.05-0.1$\arcsec$ or 5-10 AU) in both the 
aromatic and weak additional features. 
This gap could be due to the lack of the carrier of the bands in the regions very near the star or to the fact that the emission of these features disappears completely in contrast to the continuum. 
But it could also result from the presence of a ``physical hole''  
as suggested by the analysis of the ISO spectra of HD~100546 by \cite{bouwman2003}.  
Higher spatial resolution data combined with an accurate analysis of the
SED are necessary to shed light on this very interesting object. 

\item We  detect clearly the HI Pf $\gamma$ recombination line in 
practically all
 stars; the line is not spatially resolved, and  probably originates
from ionized material very close to the star,
possibly  the ionized gaseous disk or matter accreting from the inner disk onto the star.

\item We compare the observed spatial distribution of the 3.3~$\mu$m PAH feature and the adjacent continuum to the predictions of a disk model that includes 
transiently heated small grains and PAHs in addition to large grains \cite[]{habart2004a}.
The model predicts that, as observed, the 3.3~$\mu$m PAH emission feature is 
significantly broader than that of the adjacent continuum
and that 50\% of its integrated intensity comes from a radius $R<$ 30 AU.
We find a very good agreement of the predicted brightness profiles with our observations.
This gives a strong support for the physical pictures underlying our model
and show that the 3.3~\um\ PAH feature comes mostly from the inner disk regions.

\item Finally, our results
 have the interesting implication that PAHs survive photo-evaporation or coagulation processes 
even if these processes are in principle important in the inner disk,  
where the UV flux and the gas density are very high.
Future detailed comparison with theoretical calculations will certainly add important information to the studies of  the processes affecting the 
size of grains  in disks.

\end{enumerate}

\begin{table*}[h!] \caption{Parameters of the targets. Evidence for a flared disk from spectral energy
distribution modeling, infrared and millimeter interferometry observations. Presence of AIBs and silicate bands. Integrated strength of the aromatic 3.3~\um\ feature.}
\label{table_sample}
\begin{tabular}{llllllllllllll}
\hline
\hline
Object& d & Sp. Type & \Teff & \Lstar & $\chi$$^a$ & Flared & AIBs & $I_{3.3}^b$& Sil. & Group & Ref. \\  
       & [pc] &  & [K]  & [\Lsun] &    & disk & & [10$^{-14}$~W/m$^2$] & && & \\  
\hline
HD 169142 & 145 & A5Ve & 10500 & 32 &2.6 10$^7$& $\surd$  & $\surd$ &1(0.2) & - & Ib &  (1)  \\
HD 97048 &180  &B9-A0  &  10000    & 31 & 2 10$^7$     & $\surd$ &$\surd$ & 1.3(0.3)  & -& Ib & (2) \\
HD 100453& 114 & A9Ve & 7500 & 9& 1.3 10$^6$  & $\surd$  & $\surd$& 1.3(0.2) & - & Ib &  (1)  \\
HD 100546& 103 & B9Ve & 11000 & 36& 3.6 10$^7$& $\surd$  & $\surd$ &2.5(0.5) & $\surd$ & Ia& (1) \\
AB Aur & 144 & B9/A0Ve &  9750 & 47& 2.8 10$^7$ & $\surd$ & $\surd$ & $<1$$^c$ & $\surd$& Ia  & (1)\\
CQ Tau & 100 & F2 & 7500 & 5 & 7 10$^5$ & $\surd$ & ? & & ?  & I  & (3)\\
HD 142666 & 116 & A8Ve & 8500 & 11 &3 10$^6$ &  no   & $\surd$ & 0.3(0.2)& $\surd$ & II  & (1)\\
HD 163296  & 122 & A3Ve & 10500 & 30 & 2.4 10$^7$& no  & -&  & $\surd$  & II  & (1) \\
\hline
\end{tabular}

$\surd$ : detection, ?: possible detection, -: no detection. \\
$^a$ Far-ultraviolet (FUV, 6$<h\nu<$13.6 eV) flux at 10 AU from the star expressed in units of the average interstellar radiation field, $1.6~10^{-6}$ W m$^{-2}$ \cite[]{habing68}. \\
$^b$ Integrated strenght (after continuum subtraction) and uncertainty (in between brackets) estimated from the ISO spectrum of \cite{meeus2001} and \cite{vankerckhoven2002}.\\
$^c$ From \cite{brooke93}.\\
References: (1) \cite{meeus2001} and references therein;
(2) \cite{vankerckhoven2002} and references therein; (3) \cite{natta2001}.
\end{table*}

\begin{table*}[h!] \caption{Summary of the observations.}
\label{table_obs}
\begin{tabular}{llllllll}
\hline
\hline
Object & date & $r_0$ & ${\cal L}_0$ & SR $^a$ & seeing & standard stars\\
       &      &  [cm]  & [m]            & [\%]      & [arcsec] & \\
\hline
HD 169142 & 2004/03/12 & 7 & 22 & 25 & 1.5 & HR 5494\\
HD 97048$^b$ &  2003/01/11 &  10 & 15 & 40 & 1.1 & HIP 53074, HIP 20440 \\
HD 100453 & 2004/01/30   & 14 & 13 & 55 & 0.8 & HIP 062327\\
HD 100546  & 2004/01/31   & 11 & 16 & 45 & 1 & HIP 046283\\
AB Aur & 2003/11/04 & 11 & 23 & 30 & 0.75 & GJ1055\\
CQ Tau  & 2003/12/02  & 13 & 20& 45 & 0.6& HIP025337\\
HD 142666 & 2004/03/07  & 21 & 22 & 60 & 0.75 & HIP 074752 \\
HD 163296 & 2004/03/15 & 25 & 23& 60& 0.6& HIP 085389\\
\hline
\end{tabular}

$^a$ Resulting average coherent energy (after AO correction and at 2.2~\um) giving the
actual Strehl ratio characterizing the quality of our data.\\
$^b$ The data obtained for HD 97048 has been previously presented in \cite{habart2004c}.\\
\end{table*}

\begin{figure*}[htbp]
\leavevmode
\centerline{ \psfig{file=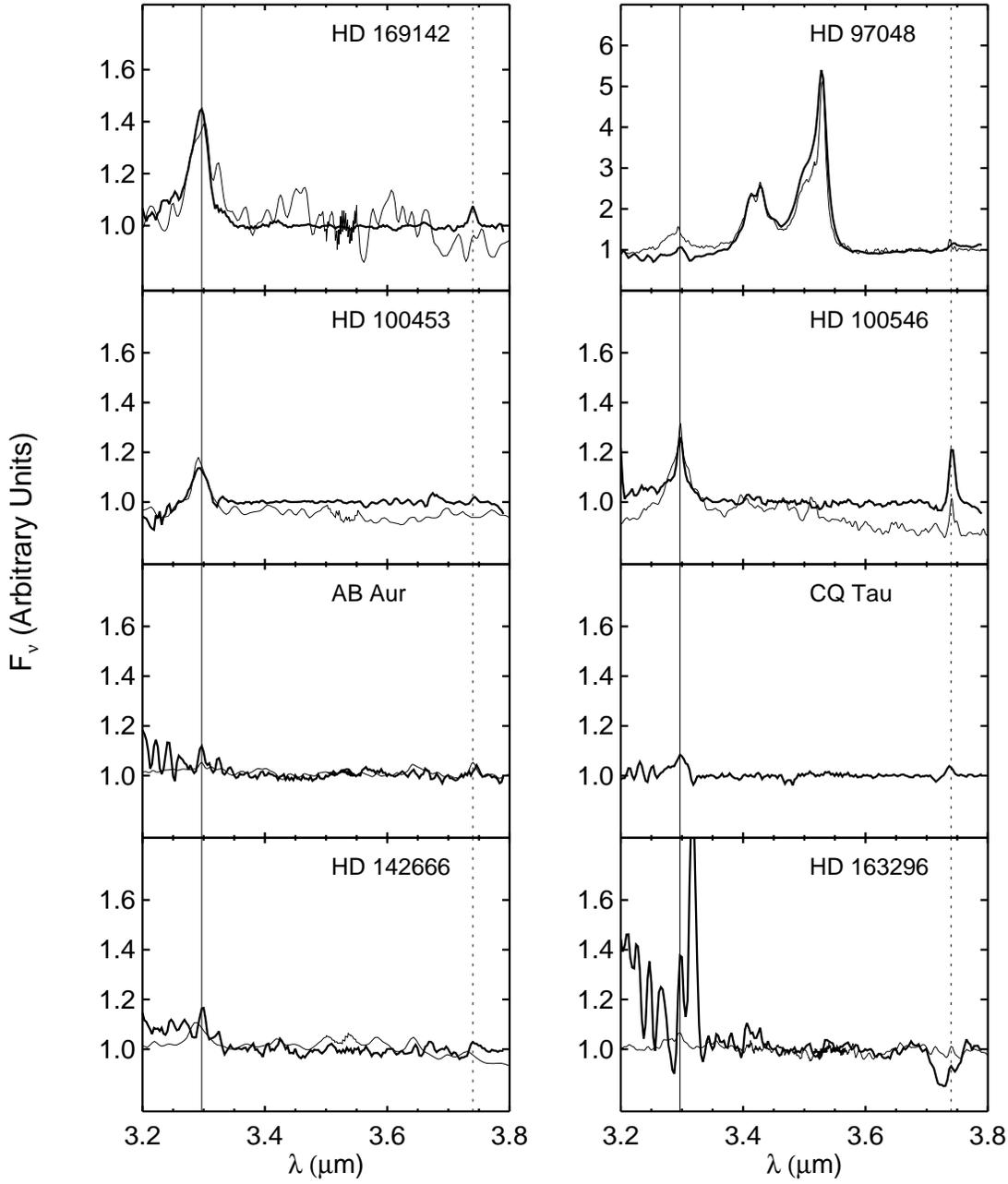,angle=0,width=16cm}}
\caption{NACO spectra obtained for distances from the star $\le$1\arcsec\ (thick solid lines). The thin lines show the ISO-SWS spectrum,
obtained with a beam of 14$\times$20\arcsec. We also show as solid vertical lines the PAH feature position and as vertical dotted lines the position of Pf~$\gamma$
line at 3.74~\um. Both NACO and ISO spectra are normalized to the continuum emission  adjacent to the 3.3~\um\ PAH feature. For HD~163296, the data were noisy and it was difficult to remove the telluric features.}
\label{spectra}
\end{figure*}

\newpage
\begin{figure*}[htbp]
\leavevmode
\centerline{ \psfig{file=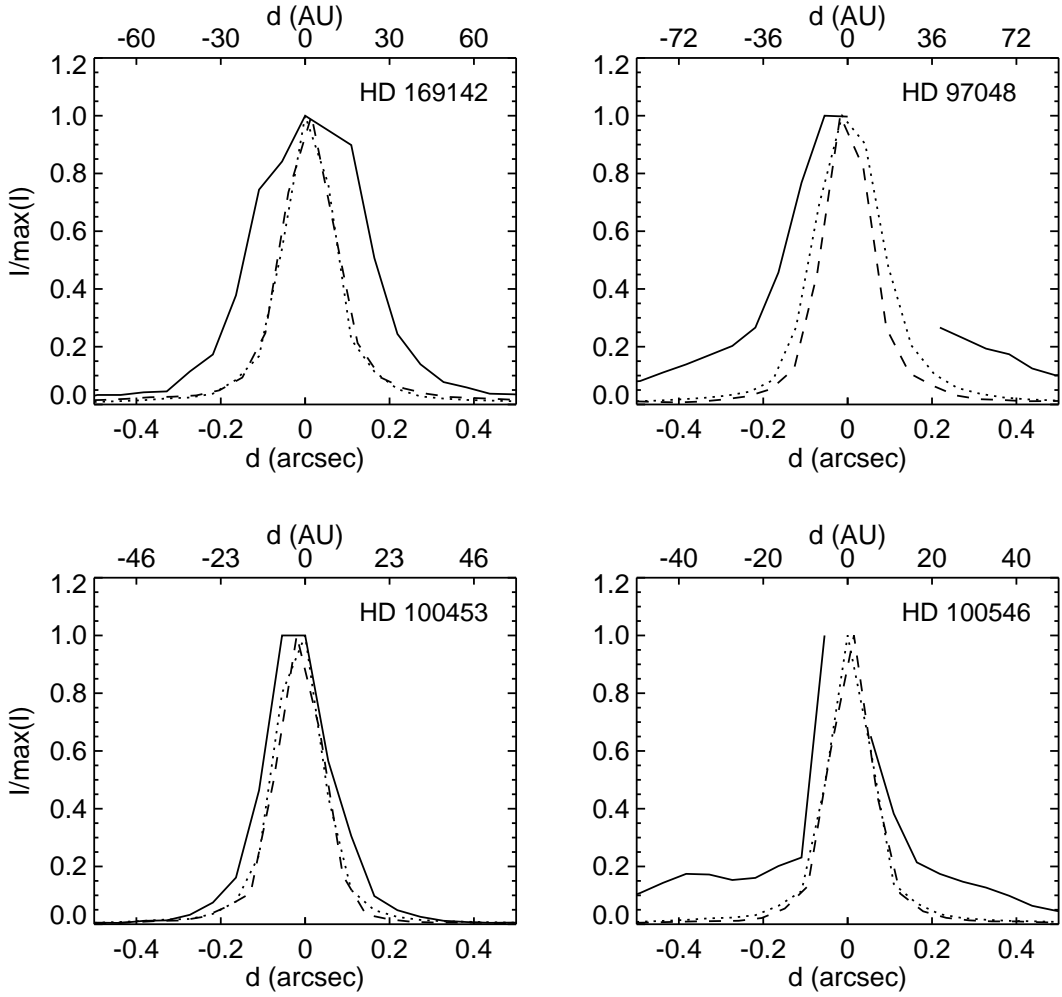,angle=0,width=16cm}}
\caption{Normalized spatial emission profiles of the 3.3~\um\ PAH feature (continuum subtracted,
          solid lines) as function of the distance from the star.  Dotted and dashed lines show the
          intensity profile of the adjacent continuum and of the PSF, respectively. 
The scale on the lower axis shows
the distance  from the star in arcsec, that on the upper axis the corresponding value in AU.
For HD~97048, on one section of the slit and for distances from the star from 0 to 0.2$\arcsec$, the NACO spectrum was noisy near the 3.3~\um\ feature and it was not possible to measure its brightness profile with sufficient accuracy.
For HD~100546, we detect the presence of a gap in the innermost region ($R\le$ 0.05-0.1$\arcsec$ or 5-10 AU); in this region, the spectrum at 3.3~\um\ is in fact completely dominated by the contribution of Pf~$\delta$ and we do not see the PAH emission (see text for more details).}
\label{profiles}
\end{figure*}

\newpage
\begin{figure*}[htbp]
\leavevmode
\centerline{ \psfig{file=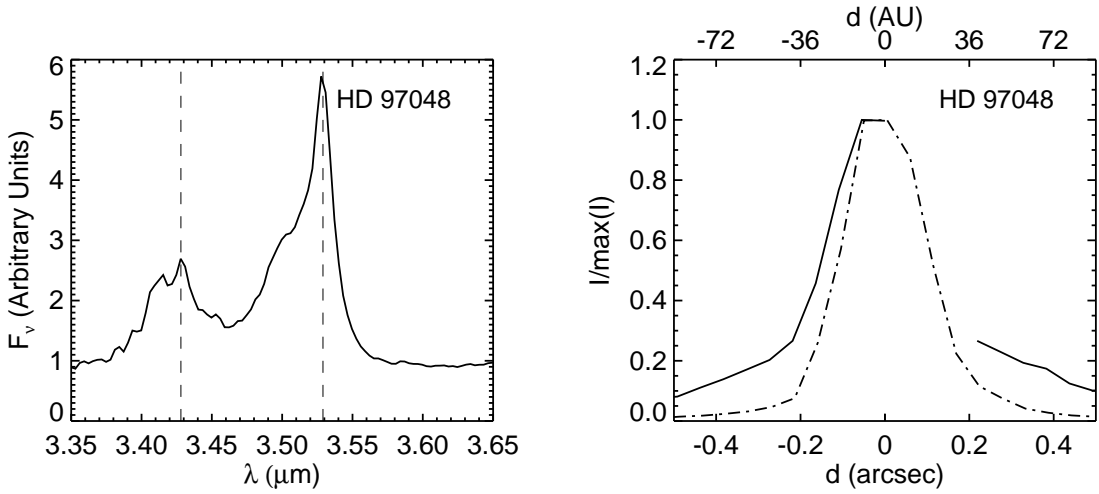,angle=0,width=16cm}}
\caption{Left panel : Spectrum of HD~97048 in the 3.35-3.65~\um\ range obtained for distances from the star $\le$1$\arcsec$. The dotted vertical lines show the strong diamond feature positions. Right panel: Normalized spatial emission profiles of the 3.3~\um\ PAH features (continuum subtracted,
          solid lines) and the 3.53~\um\ diamond feature (continuum subtracted, dotted-dashed line) as function of the distance from the star. The scale on the lower axis shows
the distance  from the star in arcsec, that on the upper axis the corresponding value in AU.}
\label{profiles_add}
\end{figure*}

\begin{figure*}[htbp]
\leavevmode
\centerline{ \psfig{file=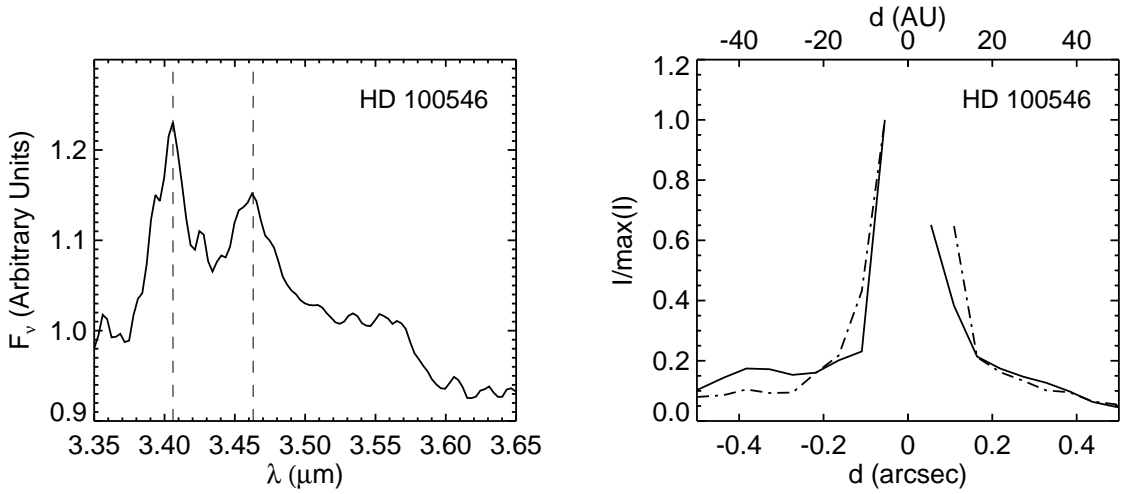,angle=0,width=16cm}}
\caption{Left panel : Spectrum of HD~100546 in the 3.35-3.65~\um\ range obtained for distances from the star going from 0.2 to 0.6$\arcsec$. The dotted vertical lines show the positions of the additional features. Right panel: Normalized spatial emission profiles of the 3.3~\um\ PAH features (continuum subtracted,
          solid lines) and the 3.4~\um\ feature (continuum subtracted, dotted-dashed line) as function of the distance from the star. The scale on the lower axis shows
the distance  from the star in arcsec, that on the upper axis the corresponding value in AU.}
\label{profiles_add_bis}
\end{figure*}

\begin{figure*}[htbp]
\leavevmode
\centerline{ \psfig{file=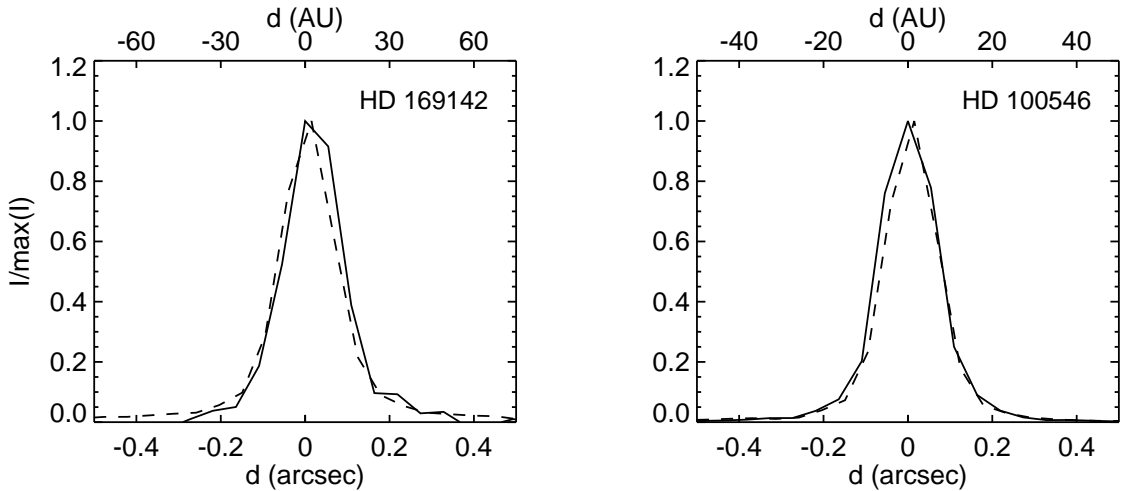,angle=0,width=16cm}}
\caption{Normalized spatial emission profiles of the HI line at 3.74~\um\ (continuum subtracted,
          solid lines) as function of the distance from the star. Dashed lines show the
          intensity profile of  the PSF. The scale on the lower axis shows
the distance from the star in arcsec, that on the upper axis the corresponding value in AU. }
\label{profiles_HI}
\end{figure*}

\newpage

\newpage
\begin{figure*}[htbp]
\leavevmode
\centerline{ \psfig{file=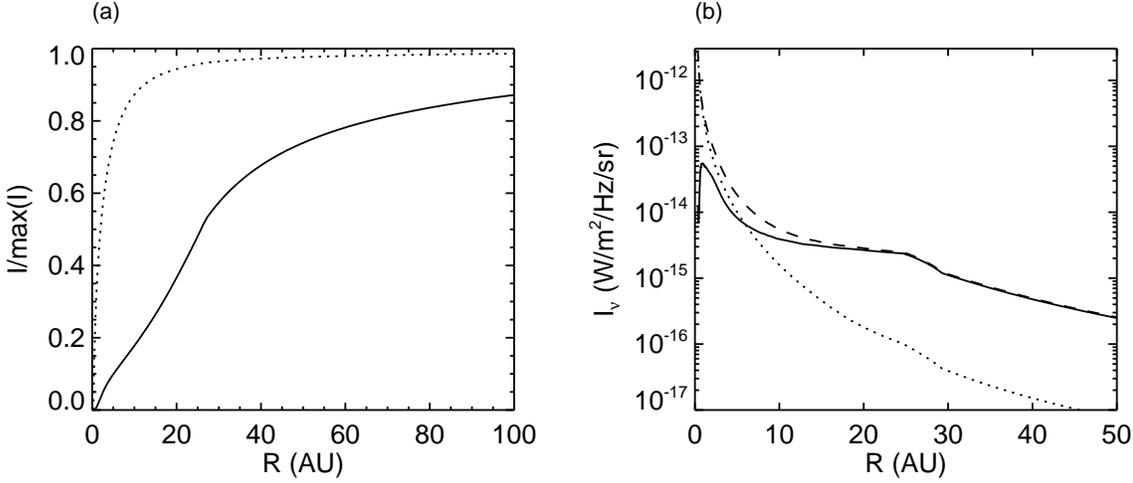,angle=0,width=16cm}}
\caption{Disk model predictions. Panel (a): Cummulative intensity of the PAHs feature (continuum subtracted, solid line) and continuum (dotted line) at 3.3~$\mu$m as function of the projected radius. Panel (b): Predicted surface brightness profiles at the peak of the 3.3~\um\ feature (no continuum subtracted,
dashed line) and in the adjacent continuum (dotted line). We also show as solid line the 3.3~\um\ feature continuum subtracted. 
  The disk is heated by a typical HAe star with effective temperature  \Tstar=10,500 K, luminosity \Lstar=32 \Lsun\
and mass \Mstar=2.4 \Msun. The disk is flared, it has a mass of 0.1 \Msun, an inner radius of 0.3 AU and an outer radius of 300 AU. The star/disk system is seen face-on at a distance of 150 pc.}
\label{profile_mod}
\end{figure*}

\newpage
\begin{figure*}[htbp]
\leavevmode
\centerline{ \psfig{file=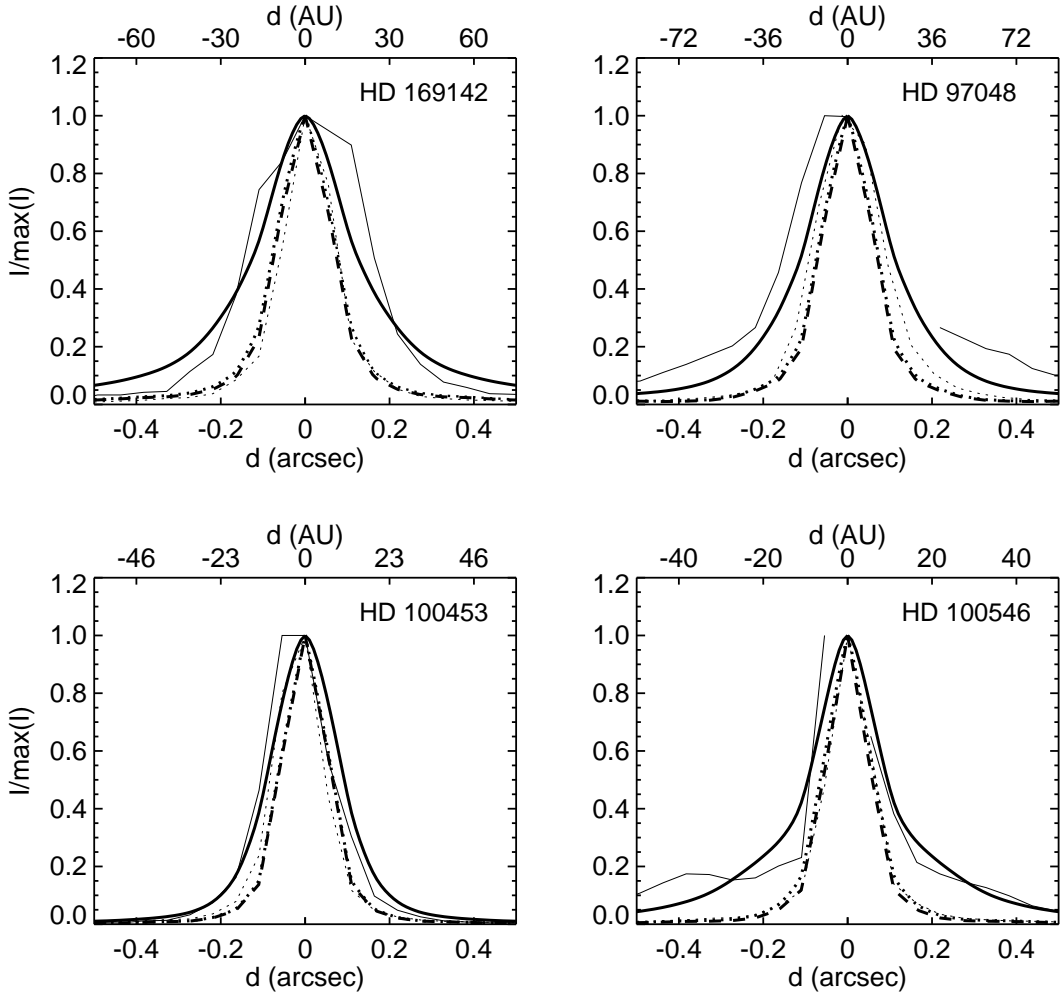,angle=0,width=16cm}}
\caption{Comparison of model predictions and observations.
Normalized spatial emission profiles of the 3.3~\um\ PAH feature (continuum subtracted,
          solid lines) and of the adjacent continuum (dotted lines) as function of the distance from the star.
The thick and thin lines show the predicted and observed spatial emission profiles, respectively.
For each object, the predicted emission profiles have been convolved with the corresponding observed PSF.
The star/disk system is assumed to be seen face-on at the distance of the object.
In each panel, the dashed line shows the normalized spatial emission profile of the PSF.
The scale on the lower axis shows
the distance from the star in arcsec, that on the upper axis the corresponding value in AU.}
\label{comparison_mod_obs}
\end{figure*}

\end{document}